\documentclass[nofigure]{pasj00}
\draft
\usepackage{graphics}
\SetRunningHead{H.Matsumoto, T.Fukue, and H.Kamaya}
{The expansion law of planetary nebulae}
\title{Interpretation of the expansion law of planetary nebulae}
\author{Hiroko Matsumoto, Tsubasa Fukue, and Hideyuki Kamaya}
\affil{%
   Department of Astronomy, School of Science, Kyoto University, Sakyo-Ku, Kyoto 606--8502}
      \email{kamaya@kusastro.kyoto-u.ac.jp}
      \Received{$\langle$reception date$\rangle$}
      \Accepted{$\langle$acception date$\rangle$}
      \Published{$\langle$publication date$\rangle$}
\begin{document}
\maketitle
\begin{abstract}

We reproduce the expansion velocity--radius
($V_{\rm{exp}}$--$R_{\rm{n}}$) relation in planetary nebulae by
considering a simple dynamical model, in order to investigate the
dynamical evolution and formation of planetary nebulae.  In our model,
the planetary nebula is formed and evolving by interaction of a fast
wind from the central star with a slow wind from its progenitor, the AGB
star.  In particular, taking account of the mass loss history of the AGB
star makes us succeed in the reproduction of the observed
$V_{\rm{exp}}$-$R_{\rm{n}}$ sequence.  As a result, 
examining the ensemble of the observational 
and theoretical evolution models of PNe, 
we find that if
the AGB star pulsates and its mass loss rate changes with time (from
$\sim 10^{-6.4}M_{\odot}$ yr$^{-1}$ to $\sim 10^{-5}M_{\odot}$
yr$^{-1}$), the model agrees with the observations.  In terms of
observation, we suggest that there are few planetary nebulae with larger
expansion velocity and smaller radius because the evolutionary
time-scale of such nebulae is so short and the size of nebulae is so
compact that it is difficult for us to observe them.

\end{abstract}
%
%
\section{Introduction}
Planetary nebulae (PNe) are the objects apart from the asymptotic giant
branch (AGB), evolving to white dwarfs. It is thought that a stellar
wind is important for the formation and evolution of PNe. To explain the
features of PNe, Kwok et al.  (1978) propose the interacting stellar
wind (ISW) theory. After that many authors have studied PNe
theoretically and observationally based on the ISW theory.
  
Observationally, before the ISW theory was presented, a trend of the
expansion velocity ($V_{\rm{exp}}$) of the gas of PNe was found (Kwok et
al. 1978), and it has been interpreted in two different ways: one is in
terms of a dynamical evolution of PN (Smith 1969; Bohuski and Smith
1974; Sabbadin and Hamzaoglu 1982; Robinson et al. 1982), another is in
terms of the evolution of a central star (Renzini 1979).
     
Based on such studies, Sabbadin et al. (1984; hereinafter SGBO) proposed
two models to interpret that trend, the ``two-wind'' model and the
``two-phase'' model.  The former means that the planetary nebula is the
region of interaction of a fast wind emitted by the nucleus with a slow
wind expelled by its progenitor.  They added the effect of the momentum
transfer from radiation field to the model suggested by Kwok et
al.(1978) and Kwok(1982).  The latter means that the planetary nebula is
a Str$\ddot{\rm{o}}$mgren sphere evolving in an expanding nebula which
results from a sudden ejection by the AGB star.  They conclude only the
``two-phase'' model agrees with the observational data because the
``two-wind'' model can not reproduce the observed mass ($M_{\rm{n}}$)
and radius ($R_{\rm{n}}$) relation.  In the two-wind model the resulting
mass of nebula is much smaller than that derived from observations (the
mass decrement problem), while it matches the
$V_{\rm{exp}}$--$R_ {\rm{n}}$ relation. The mass decrement problem
of the two-wind model is also stressed and
examined in Schmidt-Voigt \& K$\ddot{\rm{o}}$ppen
(1987a,b). 

The ``two-wind'' model had another fault. SGBO didn't take account of
the effect of shock sufficiently. 
As Marten \& Sch$\ddot{\rm{o}}$nberner (1991) 
stated, the simplification of SGBO corresponds to the fact
that the width of the shell is infinitesimally thin. 
To resolve the mass decrement problem, it is essential to
examine the width of the shell (Marten \& Sch$\ddot{\rm{o}}$nberner 1991).
Then, we confirm in this paper that
it is easy to resolve the mass decrement problem 
if the width of the shell is determined as the effect of the shock.
In this standing point, we try to study the two-wind model
to satisfy $M_{\rm{n}}$--$R_{\rm{n}}$ and 
$V_{\rm{exp}}$--$R_{\rm{n}}$ relations.
We would like to comment that
after SGBO, they developed their study on PNe furthermore
(e.g. Sabbadin et al. 2005).

We also examine the
effect of the time-dependent mass loss rate of the AGB star, then we
succeed in reproduction of the observed mass-radius relation
simultaneously with the observed $V_{\rm{exp}}$-$R_{\rm{n}}$ relation.
The importance of the mass-loss to explain the evolution of PNe
has been stressed by  Marten \& Sch$\ddot{\rm{o}}$nberner (1991).
In this paper, we insist furthermore
that our modified ``two-wind'' model reproduces
the ensemble of observational data adequately if the radial density 
profile of
AGB matter has a sudden decrement at about 0.1pc from the central star.
This decrement is the result of the time-dependent mass-loss history
of the AGB star. Thus, we reformulate the model concerning the mass-loss
history of the AGB star.

Thus, to explore the evolution and formation of PNe,
we revisit the ``two-wind" model in this paper.
In $\S$2 we describe our model. In $\S$3 the results of the calculation
compared with the observational data are presented.  In $\S$4, we
discuss the results, and summarize our study in $\S$5.

\section{Model description}

According to SGBO, there are two problems in the classical ``two-wind'' model:
(a) the resulting mass of the PN is much smaller than that determined
from observations [hereinafter problem(a)], (b) the ``two-wind'' model
can't reproduce the observed $R_{\rm{n}}$-$M_{\rm{n}}$ relation [See
Fig.6 and Fig.7 of SGBO and hereinafter problem (b)]. Then, we
reformulate the problem here.
According to Frank (1994), it is important for us to study
the evolution of PNe as a problem of radiation hydrodynamics. We admit
it. However, if we are concerning the ensemble of the evolution of PNe,
some simple models of PN evolution is useful, since the overall property
of evolution
of PNe is depicted as a combination among simple physics. Then, we try
to reconstruct the two-wind model and explore its  possibility 
as a tool to study PN evolution. 

\subsection{General description}\label{seq:general}
\begin{figure}[t]
  \scalebox{0.65}{
    \includegraphics{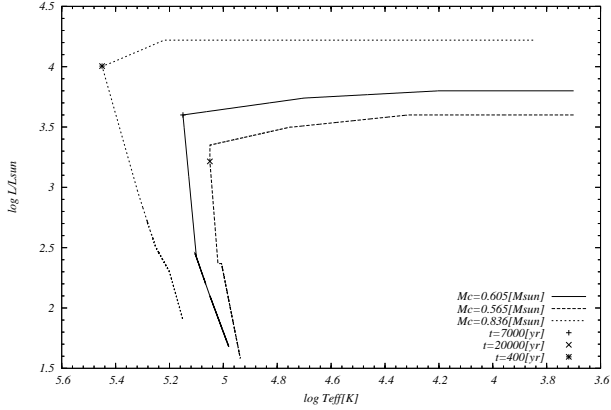}
    \label{fig:evolution}
  }
  \caption{The evolutionary tracks of central stars.  
  Horizontal axis is effective temperature, 
   and vertical axis denotes luminosity. 
In this figure, three different lines correspond different mass: 
0.836$M_{\odot}$(dotted line), 0.605$M_{\odot}$(solid line), 
0.565$M_{\odot}$(dashed line).   }
\begin{tabular}{p{7.0cm}}   
\\ \hline
\end{tabular}
\end{figure}
Our model is based on the ``two-wind'' model which is proposed by SGBO.
In the ``two-wind'' model, the authors assume that PN is the region of
interaction of a fast wind from the central star with a slow wind from
the AGB star, considering the effect of a momentum transfer from the
radiation field to the nebula motion.  We assume a shell of PN is a
shock region and the nebula evolution obeys the following equations:

\begin{eqnarray}
{dM_{\rm{n}}\over{dt}}={\dot{M_{\rm{agb}}(t)}\over{V_{\rm{agb}}}}v_{\rm{n}}
\label{eq:mass}
\end{eqnarray}
and
\begin{eqnarray}
{dv_{\rm{n}}\over{dt}}={1\over{M_{\rm{n}}}}\left[\alpha{L\over{c}}
                  -{\dot{M_{\rm{agb}}}(t)\over{V_{\rm{agb}}}}
           (v_{\rm{n}}-V_{\rm{agb}})^{2}+
           {\dot{m}\over{v}}(v-v_{\rm{n}})^{2}\right]
\label{eq:motion} 
\end{eqnarray}
where $\dot{M_{\rm{agb}}}(t)$ and $V_{\rm{agb}}$ are the mass loss rate
and the velocity of the wind from the AGB star; $L$, $\dot{m}$, and $v$
are the luminosity, the mass loss rate, and the velocity of the wind
from the central star; $M_{\rm{n}}$, $R_{\rm{n}}$, and $v_{\rm{n}}$ are
the mass, the radius, and the expansion velocity of the PN; $c$ is the
light velocity; $\alpha$ is the efficiency factor of the transfered
momentum.

Eq.(\ref{eq:mass}) means that the mass of PN is consisted of the total
matter included in the radius $R_{\rm{n}}$ thanks to the shock.
$\dot{M_{\rm{agb}}}(t)$ denotes the mass loss rate measured at the
position of the expanding shell, and then it is represented of time
explicitly.

In Eq.(\ref{eq:motion}), the first term in the right hand side represents 
the contribution of the transfer of momentum from the radiation field to the 
nebula, the second term is the interaction with the AGB slow wind, 
and the third term corresponds to the interaction with the fast wind from the
central star.

\subsection{Models of the central star}\label{seq:central}
Adopting Eqs.(\ref{eq:mass}) and (\ref{eq:motion}), we take into account
the evolution of luminosity $L$ and the mass-loss $\dot{m}$ of the
central star more realistically than SGBO adopted, while those evolution
models are simplified as presented below.

We assume that the mass of the central star is $0.605M_{\odot}$ because
most nuclei of PNe are about $0.6M_{\odot}$. Then we approximate the
evolutionary track proposed by Bl$\ddot{\rm{o}}$cker (1995b) shown in
Fig.1.  Furthermore, we compute the mass-loss rate shown in Fig.2,
according to Martin \& Sch$\ddot{\rm{o}}$nberner (1991).
\begin{figure}[t]
  \scalebox{0.65}{
    \includegraphics{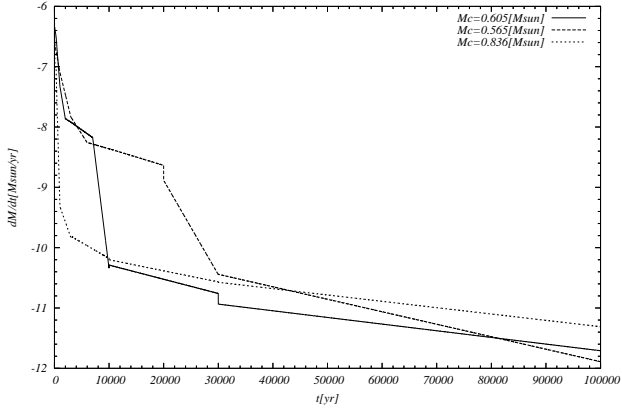}
    \label{fig:mass-loss}
   }
  \caption{Mass-loss rate of a central star.  
Horizontal axis is evolutionary time [yr] and vertical axis is mass loss rate
[$M_{\odot}$/yr]. Each of three lines corresponds to different central
 mass: dotted line 
corresponds 0.836$M_{\odot}$, solid line is 0.605$M_{\odot}$, dashed line is 
0.565$M_{\odot}$. }
\begin{tabular}{p{7.0cm}}   
\\ \hline
\end{tabular}
\end{figure}

\subsection{Initial Conditions}
As stated, SGBO insist there are two problems (mass decrement problem
and $R_{\rm{n}}$-$M_{\rm{n}}$ relation problem) for the ``two-wind''
model.  In order to solve these problems, we adopt Eq.(1) because the
shock condition has to be satisfied there and the variation of the
mass-loss rate of the AGB star to the ``two-wind'' model should be
considered.  Problem (a) is solved easily by only considering the shock
condition.  The previous model did not count the total mass reasonably.

Following Bl$\ddot{\rm{o}}$cker (1995a), we adopt the time-depending 
simplified mass-loss rate 
from the AGB star described in Fig.3 to reproduce the 
$R_{\rm{n}}$-$M_{\rm{n}}$ relation.
\begin{figure}[t]
   \scalebox{0.65}{
    \includegraphics{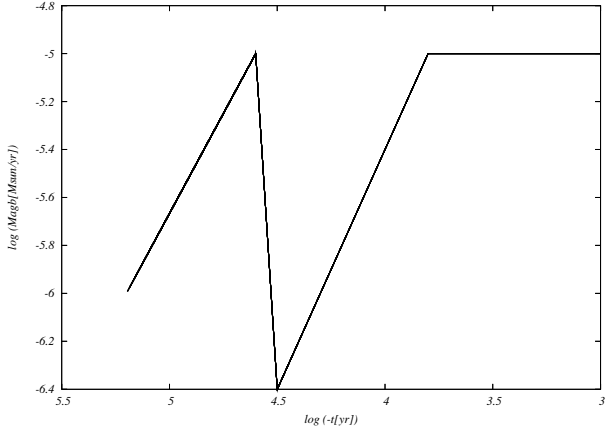}
    \label{fig:Magb}
   }
  \caption{The mass-loss rate of the AGB star. 
Horizontal axis denotes time before a star enters PN phase, and vertical one is 
mass loss rate of the AGB star.  } 
\begin{tabular}{p{7.0cm}}   
\\ \hline
\end{tabular}
\end{figure} 
In Fig.3, the horizontal axis represents the time before the star
enters post-AGB phase, i.e. the right edge of it is 1000 yr before
the post-AGB phase and the 
left edge corresponds $10^{5.5}$yr before it.
For $-t<1000$, $\dot{M_{\rm{agb}}}$ 
decreases slowly and connects $\dot{m}$ in Fig.2.

On these bases, we set, as initial conditions,
$R_{\rm{n}}$=0.01pc, $M_{\rm{n}}$=$10^{-2.5}M_{\odot}$, 
$v_{\rm{n}}$=$V_{\rm{agb}}$+$C_{\rm{s}}$ ($C_{\rm{s}}$: the sound 
speed in pre-shock matter) at $t$=0, and 
$V_{\rm{agb}}$=10 km s$^{-1}$ which is kept during all the evolutionary time.
These are chosen simply since PN emerges as a shocked layer.
{}From $t$=0 to $t$=100000 yr, we calculate some models 
with different wind velocities from the central star: 500 km s$^{-1}$ 
$\leq v\leq 
$10000 km s$^{-1}$  at the intervals of 500 km s$^{-1}$.
The initial set of the parameters 
of $R_{\rm{n}}$ and $M_{\rm{n}}$ are very consistent to
those expected in a model calculation by Mellema (1994). 
According to the estimates of the H$_\alpha$ surface brightness 
in Fig.10 and the ionized mass in Fig.19 of Mellema (1994), 
when the PN shell is formed and fully ionized, the radius
of the shell is less than 0.05 pc and its mass
is less than $10^{-2}$ $M_\odot$. Thus, we confirm
that we choose appropriate values as the initial conditions
from the study of Mellema.

\subsection{Observational sample selection}
In our analysis, we use the [OIII] data listed in SGBO (See 
their table.1 and Fig.1). We adopt this classical set of data
because number of sample sets, in which both the size and
expansion velocity are observed directly, does not increase 
even in recent years very much.
Indeed, for our purpose, we must remove some objects whose expansion 
velocities are uncertain (small points in Fig.1 of SGBO).

According to Sch$\ddot{\rm{o}}$nberner, Jacob, \& Steffen (2005),
Sch$\ddot{\rm{o}}$nberner, et al. (2005), and so on, 
not only
there are different sizes correspond to different components, optical
imaging usually measures the size of the ionization front, not the
actual size of the shell. Their statement is important if we 
are concerning the detailed structure of PNe. Fortunately, however,
we are interested in the ensemble of the evolution of PNe, and
do not examine the detailed structure of PNe. Furthermore,
the shell can be almost ionized after about 1000 years 
(e.g. Perinotto et al. 2004).
Then, we expect that as long as the overall trend of PNe
is concerned, our simple approach is still useful.

\section{Results}\label{sec:results}
\subsection{Characteristic cases}
The results of our calculation in the case of $C_{\rm{s}}$=5km s$^{-1}$ and 
$\alpha$=0.5 
are 
shown in Fig.4, Fig.5, and Fig.6.
\begin{figure}[htb]
   \scalebox{0.65}{
    \includegraphics{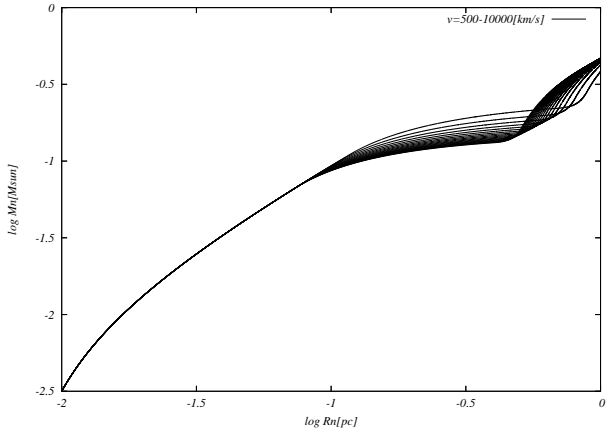}
    \label{fig:log}
   }
  \caption{Rn-Mn relation. 
Horizontal axis is radius in the unit of pc, 
vertical one is mass in the unit of solar mass.
The bottom line is $v=500$ km s$^{-1}$, and the top line is $v=10000$ km 
s$^{-1}$. }
\begin{tabular}{p{7.0cm}}   
\\ \hline
\end{tabular}
\end{figure} 
\begin{figure}[htb]
   \scalebox{0.65}{
    \includegraphics{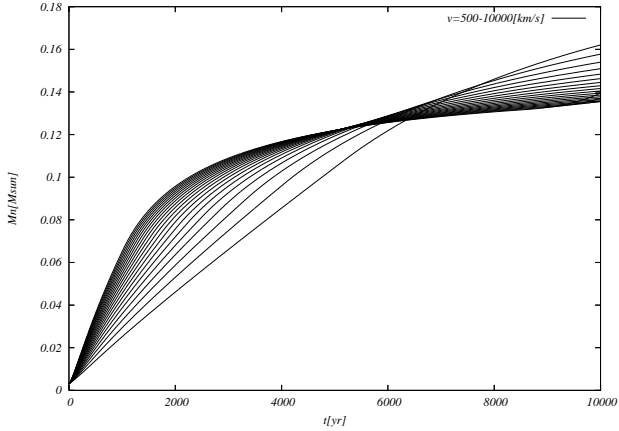}
    \label{fig:resulting mass}
   }
  \caption{Resulting mass. 
Horizontal axis corresponds evolutionary time [yr], vertical axis is mass
[$M_{\odot}$]. The bottom line is in the case of $v$=500 km s$^{-1}$, 
and the top one 
is $v$=10000 km s$^{-1}$. } 
\begin{tabular}{p{7.0cm}}   
\\ \hline
\end{tabular}
\end{figure} 
\begin{figure}[htb]
   \scalebox{0.65}{
    \includegraphics{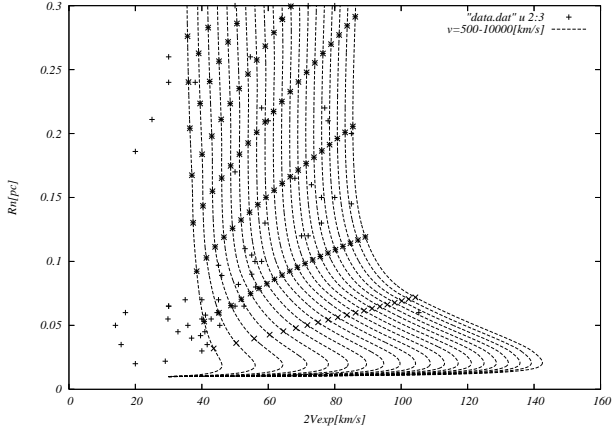}
    \label{fig:Vexp}
   }
  \caption{2Vexp--Rn relations. 
Horizontal axis denotes expansion velocity [km s$^{-1}$], vertical one is radius of PN
[pc]. The most left dashed line corresponds $v$=500 [km s$^{-1}$], 
the most right one is $v$=10000 [km s$^{-1}$]. 
Plus sign represents observational data, 
cross corresponds the point which evolutionary time is $t$=1000 yr, and
asterisks 
are marked from $t$=2000 yr at the intervals of 2000 yr. }
\begin{tabular}{p{7.0cm}}   
\\ \hline
\end{tabular}
\end{figure} 

Fig.4 represents $R_{\rm{n}}$-$M_{\rm{n}}$ relation of our calculation.
{}From this figure, because of the time-depending AGB mass-loss rate, the
increasing rate of $M_{\rm{n}}$ decreases after PN has size of
$R_{\rm{n}}\sim $0.1pc, then the increase rate turns to be small.  Thus,
we can succeed in the reproduction of the observed tendency, and resolve
the problem (b).

Around the right top of Fig.4, $M_{\rm{n}}$ increases rapidly again.
This is because the pulsation of the AGB star finishes (See Fig.3, it
corresponds $log(-t)\sim 4.5$).  However it doesn't affect the value of
expansion velocity up to $R_{\rm{n}}$=0.3 pc.  It influences only the
resulting mass of PN.  Fig.5 is time-mass plane. The bottom line is the
model $v$=500 km s$^{-1}$, the top is the model $v$=10000 km s$^{-1}$.
{}From our simple models, it takes over 6000 yr that the radius of model
nebula reaches 0.3 pc (See Fig.6), and the resulting mass is over
0.1$M_{\odot}$ when $R_{\rm n} \sim 0.3$ pc, thus we can solve the
problem (a) in the same time.

In Fig.6, the relation of $2V_{\rm{exp}}$-$R_{\rm{n}}$ is presented as
one of main results in our study.  We can almost re-product the
distribution of data points by changing only the fast wind speed. The
most left curve is the model $v=$500 km s$^{-1}$ and the most right one
corresponds $v=$10000 km s$^{-1}$.  From Eq.(\ref{eq:motion}), if $v$
has large value, the shell of PN is accelerated strongly in the early
phase then has a large expansion velocity. The expansion velocity is
decelerated as the PN evolutes, because of the decrease of the
luminosity and mass loss rate of the central star.  We also over-plot
the evolutionary time as asterisks from $t$=2000 yr at steps of 2000
yr. One can comprehend from it that most of data points exist in the
region of $t$=2000 yr-4000 yr. Star which mass is 0.605$M_{\odot}$
enters the realm of white dwarf at $t\sim$ 7000 yr (See Fig.1), so PN
phase must be before 7000 yr. Our results are consistent in this
meaning.

\subsection{Parameter dependence}
Setting the fast wind velocity $v$=1000 km s$^{-1}$, 
we inquire into the effects of 
other parameters: the sound speed ($C_{\rm{s}}$), the 
transfer efficiency ($\alpha$), and 
the slow wind velocity ($V_{\rm{agb}}$). The 
effect of other parameters to the results is essentially the same as 
examined in SGBO.
\begin{figure}[t]
   \scalebox{0.65}{
    \includegraphics{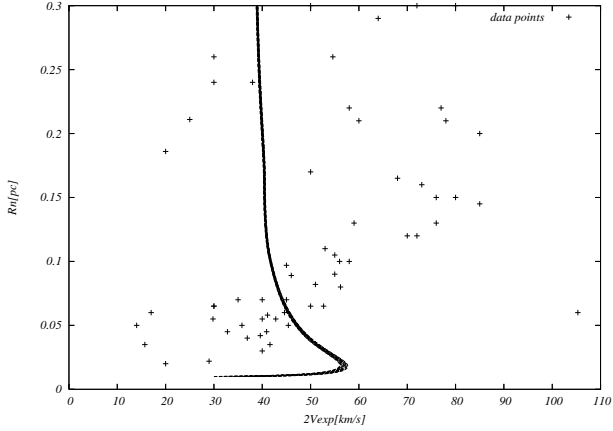}
    \label{fig:Cs}
   }
  \caption{2Vexp-Rn:Cs=5-10 km s$^{-1}$. 
Horizontal axis is expansion velocity [km s$^{-1}$], 
and vertical one is radius [pc]. 
 
Plus sign represents observational data. }
\begin{tabular}{p{7.0cm}}   
\\ \hline
\end{tabular}
\end{figure} 

\subsubsection{Sound speed}
Fig.7 shows the results of different sound speed in the pre-shock matter: 
$C_{\rm{s}}$=5-10 km s$^{-1}$ at the intervals of 0.5km s$^{-1}$. 
Even if $C_{\rm{s}}$ changes, i.e. 
the temperature in the pre-shock matter changes,
$V_{\rm{exp}}$-$R_{\rm{n}}$ relation is not so different. 
The variation of $C_{\rm{s}}$ affects 
only $R_{\rm{n}}$-$M_{\rm{n}}$ relation, as seen in Fig.8. 
It is because we set that 
the mass of PN increases along Eq.(\ref{eq:mass}) only as long as the shock 
occurs. 
The 
larger the sound speed is, the shorter the period during which
the shock is occurring is and 
the smaller the final mass of PN is.
If the shock doesn't occur, the mass of PN has a constant value
in our model.
\begin{figure}[t]
   \scalebox{0.65}{
    \includegraphics{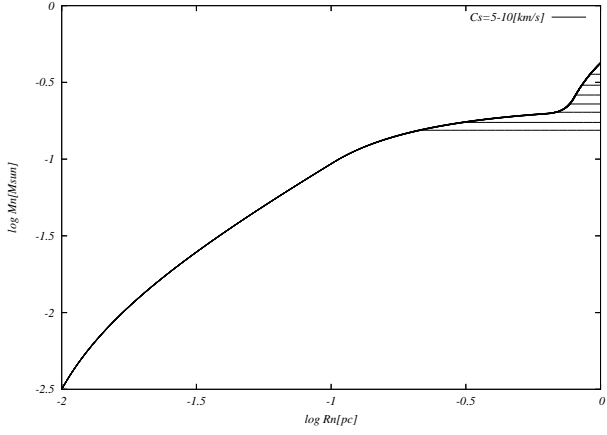}
    \label{fig:Cs_log}
   }
  \caption{log Rn-log Mn:Cs=5-10 km s$^{-1}$. 
Horizontal axis corresponds radius, and
vertical axis is mass. The bottom line is 
in the case of $C_{\rm{s}}$=10 km s$^{-1}$. }
 \begin{tabular}{p{7.0cm}}   
\\ \hline
\end{tabular}
\end{figure}
  
\subsubsection{Transfer efficiency}
The sequences of different transfer efficiency are shown in Fig.9 and Fig.10.
\begin{figure}[t]
   \scalebox{0.65}{
    \includegraphics{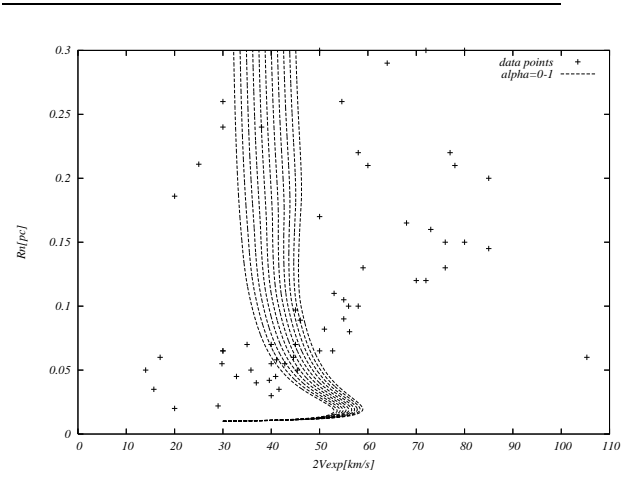}
    \label{fig:alpha}
   }
  \caption{2Vexp-Rn:alpha=0-1. 
Horizontal axis is expansion velocity [km s$^{-1}$], 
and vertical axis denotes radius [pc]. 
The most left dashed line is $\alpha$=0, the most right one is $\alpha$=1, 
and plus sign 
means observational data. } 
\begin{tabular}{p{7.0cm}}   
\\ \hline
\end{tabular}
\end{figure}
\begin{figure}[t]
   \scalebox{0.65}{
    \includegraphics{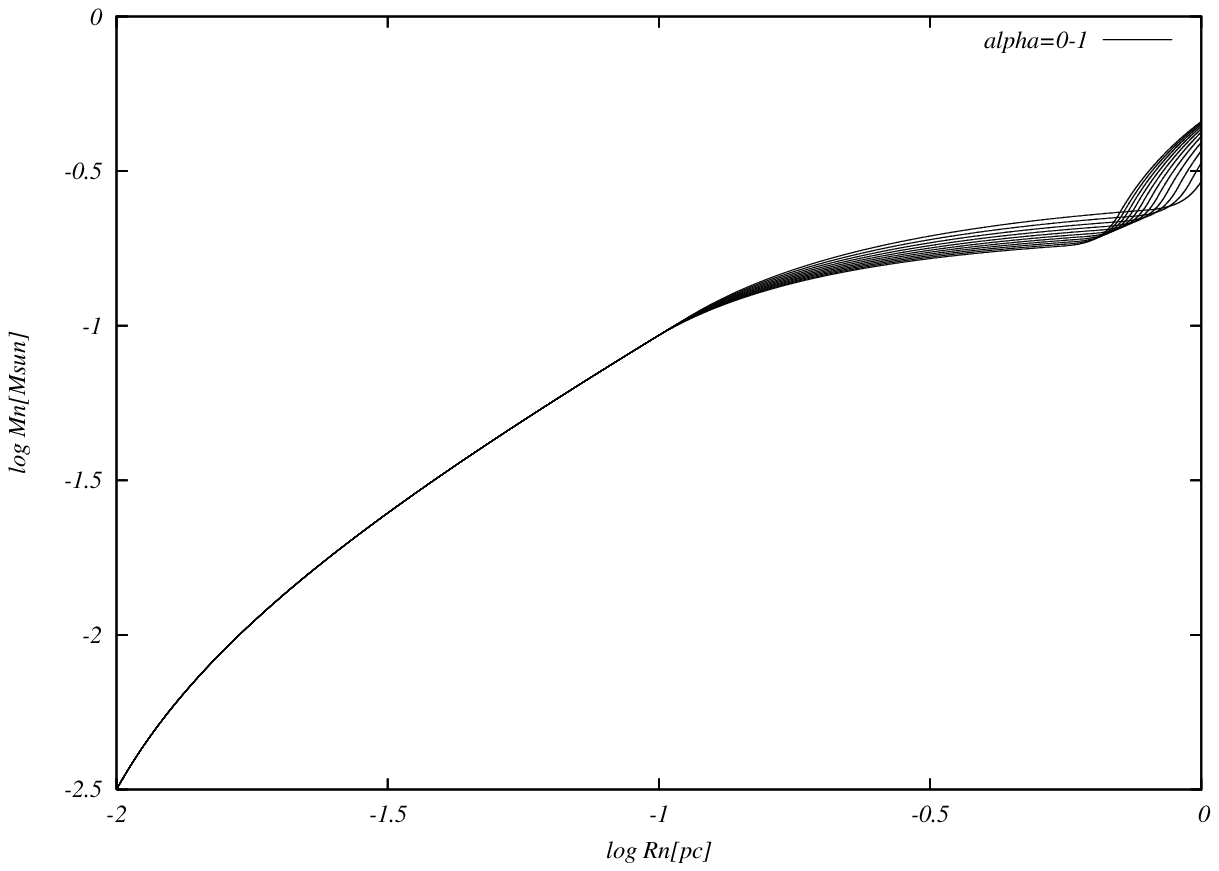}
    \label{fig:alpha_log}
   }
  \caption{log Rn-log Mn:alpha=0-1. 
Horizontal axis is radius, and vertical axis is mass. The bottom line is in the 
case of $C_{\rm{s}}$=0. }
\begin{tabular}{p{7.0cm}}   
\\ \hline
\end{tabular}
\end{figure}
We vary $\alpha$=0-1 at the steps of 0.1. As $\alpha$ increases, the 
influence on the evolution of PN 
due to the variation of central star's luminosity becomes large.  

\subsubsection{Slow wind velocity}
The $2V_{\rm{exp}}$-$R_{\rm{n}}$ relations are drawn in Fig.11, and the
$R_{\rm{n}}$-$M_{\rm{n}}$ relations are in Fig.12 when we change the
slow wind velocity from 5km s$^{-1}$ to 24km s$^{-1}$ at the intervals
of 1km s$^{-1}$.  In Fig.11, the most left line corresponds
$V_{\rm{agb}}$=5km s$^{-1}$ and the most right one does
$V_{\rm{agb}}$=24km s$^{-1}$. Also in Fig.12, the top curve is in the
case of $V_{\rm{agb}}$=5km s$^{-1}$ and the bottom is
$V_{\rm{agb}}$=24km s$^{-1}$.  If we allow a turning point, at which the
increasing rate of mass turns to be small, to have uncertainty of a
factor 3 in log ($R_{\rm{n}}$)-log ($M_{\rm{n}}$) plane, ones can find
that the difference of AGB wind velocity also explains the distribution
of observed data well.  However, AGB wind velocity is not so large
($<$30 km s$^{-1}$) that the models varying only $V_{\rm agb}$ does not
cover the region with large expansion velocity.  From this point of
view, we conclude the central star's wind velocity is more proper
parameter which explains 2$V_{\rm{exp}}$-$R_{\rm{n}}$ relation.
\begin{figure}[t]
   \scalebox{0.65}{
    \includegraphics{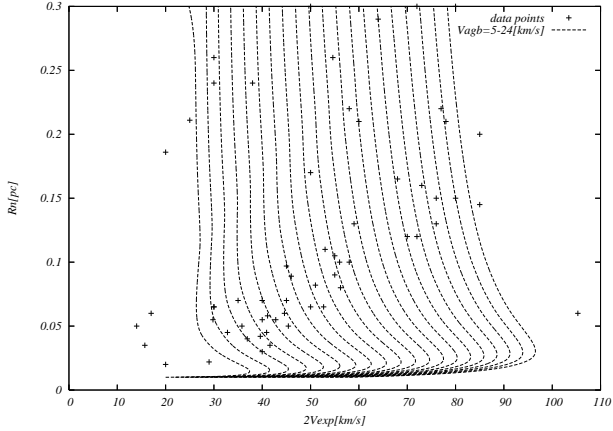}
    \label{fig:alpha}
   }
  \caption{2Vexp-Rn:Vagb=5-24km s$^{-1}$. 
Horizontal axis denotes expansion velocity [km s$^{-1}$] and vertical axis is 
radius [pc]. Plus sign corresponds observed data, 
the most left dashed line is in the 
case of $V_{\rm{agb}}$=5km s$^{-1}$, and the most right one is the model with
$V_{\rm{agb}}$=24km s$^{-1}$. } 
\begin{tabular}{p{7.0cm}}   
\\ \hline
\end{tabular}
\end{figure}
\begin{figure}[t]
   \scalebox{0.65}{
    \includegraphics{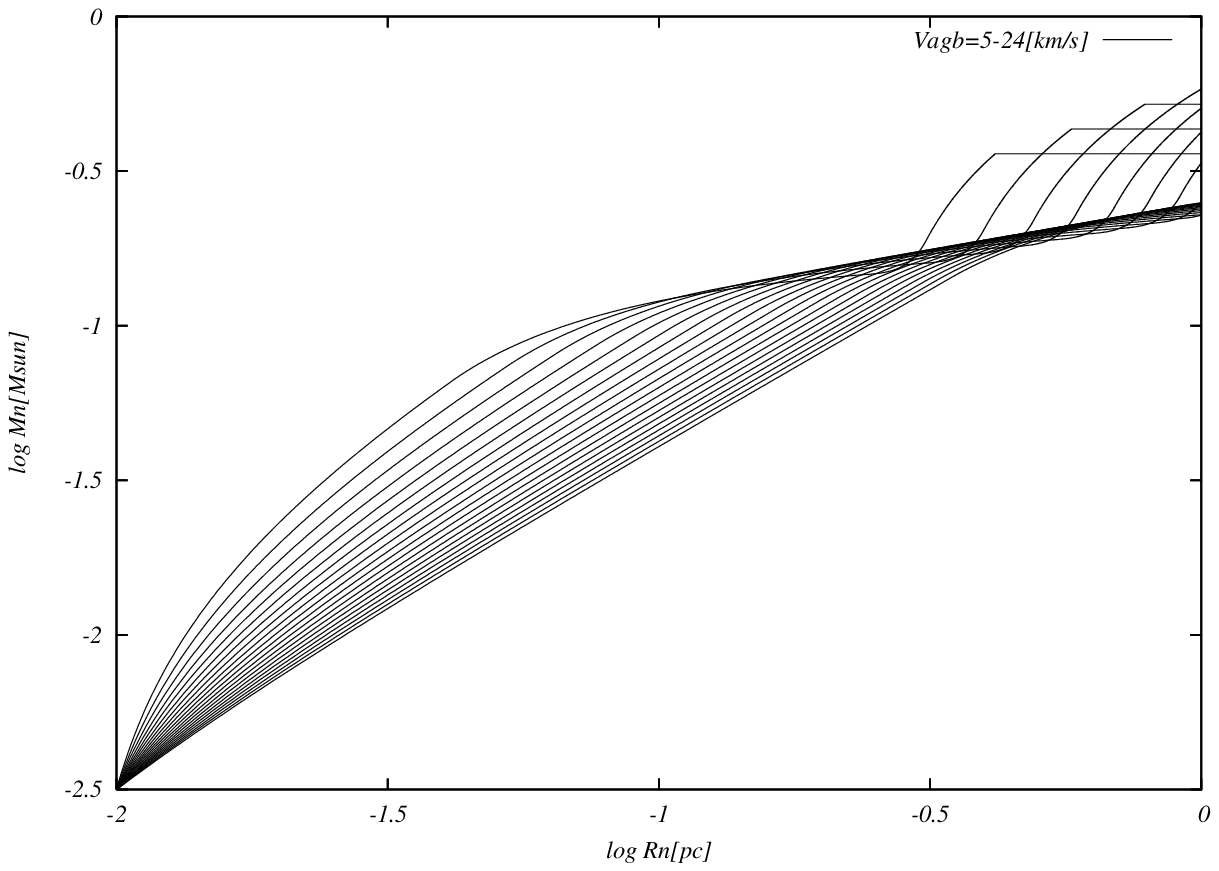}
    \label{fig:Vagb_log}
   }
  \caption{log Rn-log Mn:Vagb=5-24km s$^{-1}$. 
Horizontal axis is radius and vertical one is mass. The top line is 
$V_{\rm{agb}}$=5km s$^{-1}$ and the bottom is $V_{\rm{agb}}$=24km s$^{-1}$.}
\begin{tabular}{p{7.0cm}}   
\\ \hline
\end{tabular}
\end{figure}
    
\section{Discussion}
In this section, we examine our model compared with 
the basic model basing on SGBO and 
discuss the physics of PN evolution.

\subsection{Difference from a basic model}\label{sec:difference}
As already mentioned, there are two problems 
in the classical two-wind model by SGBO: 
(a) the decrement of resulting mass, and (b) the discrepancy in the 
$R_{\rm{n}}$-$M_{\rm{n}}$ relation between observations and the model.
Although SGBO's model explains the $2V_{\rm{exp}}$-$R_{\rm{n}}$ relation, 
the authors abandon their idea because of the problems (a) and (b). 
On the other hand, 
we take account of the effect of shock and time-depending mass loss rate of 
the AGB star, then the problems (a) and (b) are solved easily. 
{}From this point, we convince for the evolution of PN to consider the 
influence of shock and variation of mass loss rate of the AGB star. 
That is, the AGB star is pulsating so its mass loss rate varies with 
time during AGB phase, and during the PN phase the fast wind speed is much 
larger than the slow wind speed to be a shock: $v \gg V_{\rm{agb}}$. 
  
\subsection{Physical origin of the Vexp-Rn relation}
\begin{figure}[htbp]
   \scalebox{0.65}{
    \includegraphics{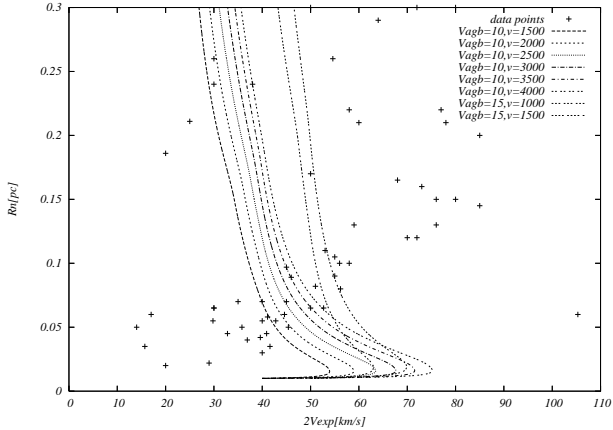}
    \label{fig:2e-5}
   }
  \caption{2Vexp-Rn:$\dot{M_{\rm{agb}}}$=2$\times10^{-5}M_{\odot}$ yr$^{-1}$. 
Horizontal axis is expansion velocity [km s$^{-1}$], and
vertical axis is radius [pc]. 
Plus sign means the data. $\dot{M_{\rm{agb}}}$=2$\times10^{-5}M_{\odot}$ yr$^{-1}$, 
$\alpha$=0.5, $C_{\rm{s}}$=10 km s$^{-1}$ and other parameter's value 
are written in the figure.  }
\begin{tabular}{p{7.0cm}}   
\\ \hline
\end{tabular}
\end{figure}
\begin{figure}[htbp]
   \scalebox{0.65}{
    \includegraphics{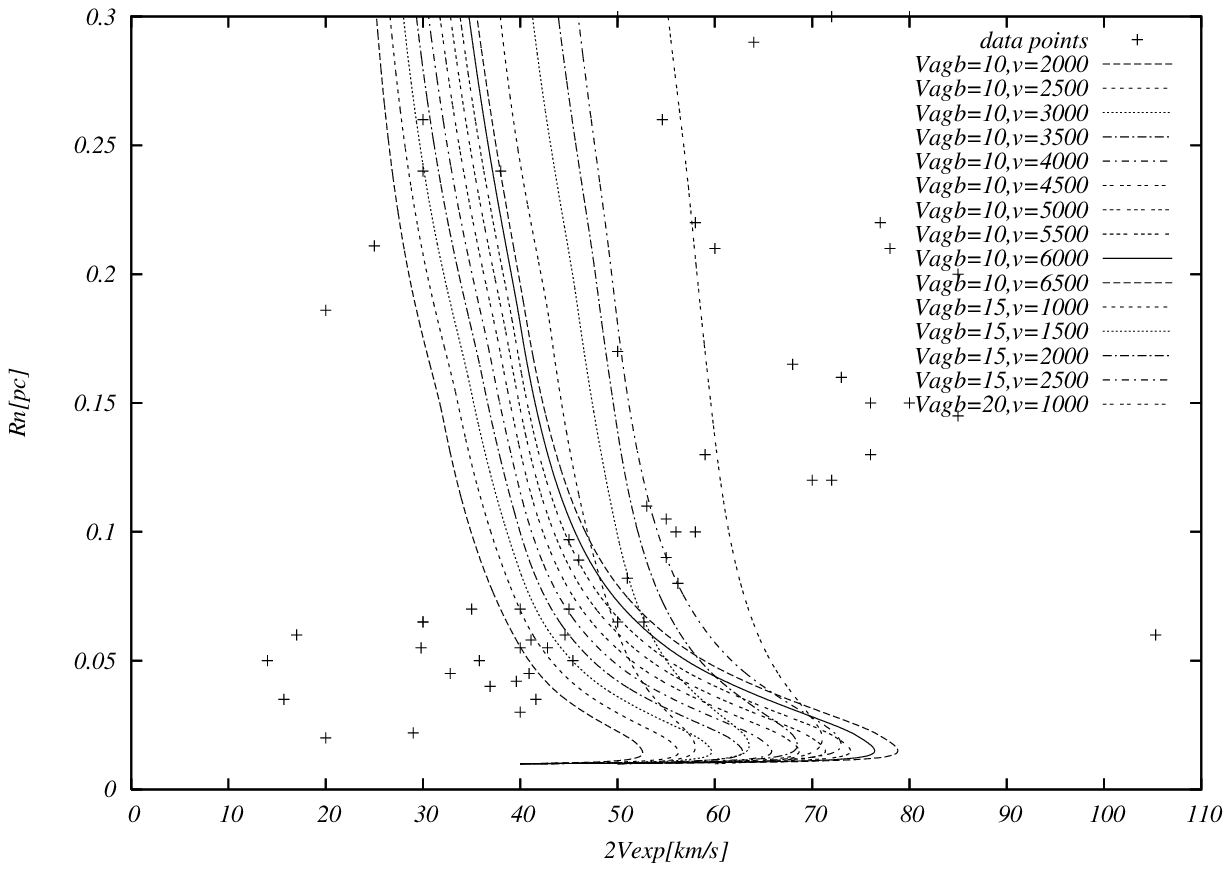}
    \label{fig:3e-5}
   }
  \caption{2Vexp-Rn:$\dot{M_{\rm{agb}}}$=3$\times10^{-5}M_{\odot}$ yr$^{-1}$.
Same as Fig.13 but in this figure 
$\dot{M_{\rm{agb}}}$=3$\times10^{-5}M_{\odot}$ yr$^{-1}$.} 
\begin{tabular}{p{7.0cm}}   
\\ \hline
\end{tabular}
\end{figure}
\begin{figure}[htbp]
   \scalebox{0.65}{
    \includegraphics{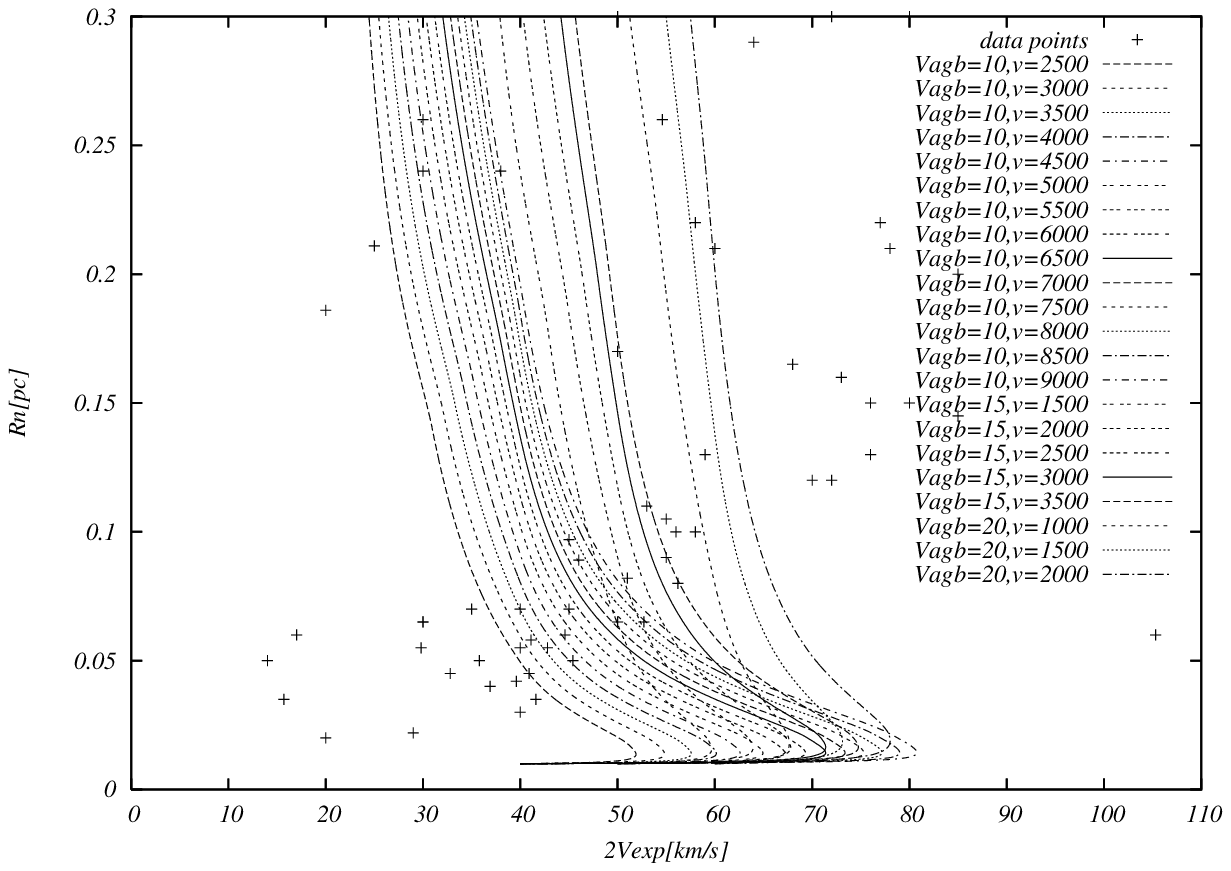}
    \label{fig:4e-5}
   }
  \caption{2Vexp-Rn:$\dot{M_{\rm{agb}}}$=4$\times10^{-5}M_{\odot}$ yr$^{-1}$.
 Same as Fig.13 but in this figure 
$\dot{M_{\rm{agb}}}$=4$\times10^{-5}M_{\odot}$ yr$^{-1}$.}
\begin{tabular}{p{7.0cm}}   
\\ \hline
\end{tabular}
\end{figure}
\begin{figure}[htbp]
   \scalebox{0.65}{
    \includegraphics{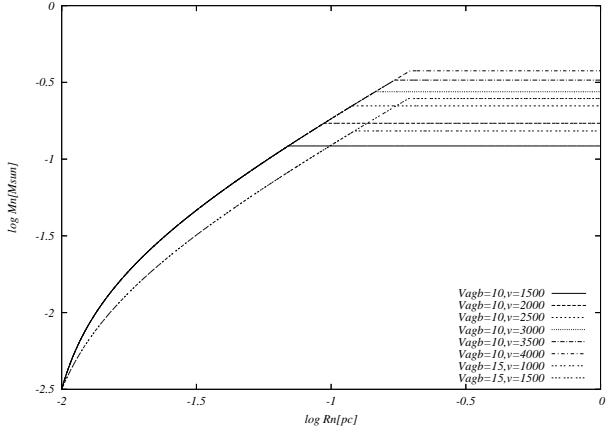}
    \label{fig:Vagb_log}
   }
  \caption
  {log Rn-log Mn:$\dot{M_{\rm{agb}}}$=2$\times10^{-5}M_{\odot}$ yr$^{-1}$. 
Horizontal axis denotes radius and 
vertical axis is mass. $\dot{M_{\rm{agb}}}$=2$\times10^{-5}M_{\odot}$ yr$^{-1}$, 
$\alpha$=0.5, $C_{\rm{s}}$=10 km s$^{-1}$ and other parameter's value 
are written in the figure. }
\begin{tabular}{p{7.0cm}}   
\\ \hline
\end{tabular}    
\end{figure}
\begin{figure}[htbp]
   \scalebox{0.65}{
    \includegraphics{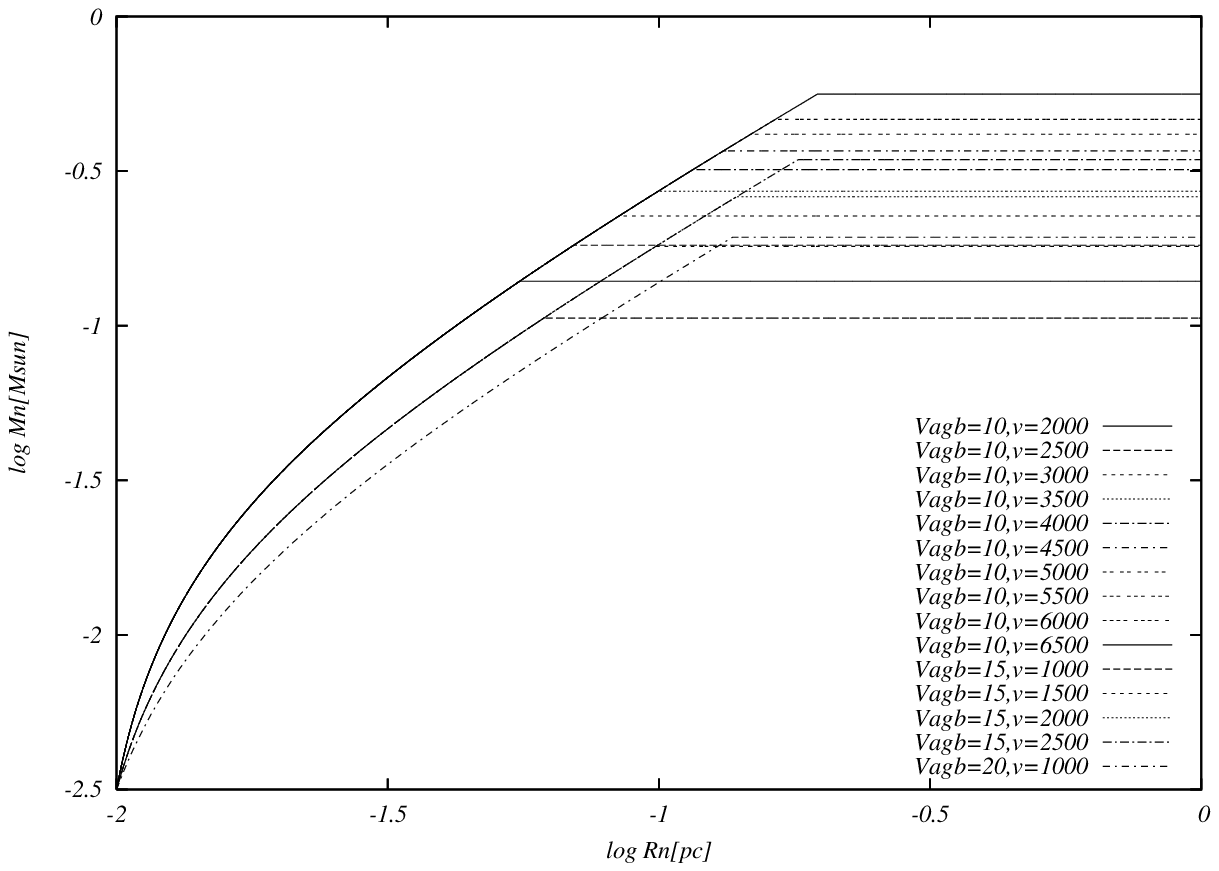}
    \label{fig:3e-5_log}
   }
  \caption{log Rn-log Mn:$\dot{M_{\rm{agb}}}$=3$\times10^{-5}M_{\odot}$ yr$^{-1}$. 
Same as Fig.16 but $\dot{M_{\rm{agb}}}$=3$\times10^{-5}M_{\odot}$ yr$^{-1}$.}
\begin{tabular}{p{7.0cm}}   
\\ \hline
\end{tabular}
\end{figure}
Based on Sect.\ref{sec:difference}, we discuss physical background of PN
evolution in this subsection.  In our model to explain the
$2V_{\rm{exp}}$-$R_{\rm{n}}$ relation, the greatest point is considering
the time-depending mass loss rate of the AGB star.  We also calculate the
models which $\dot{M_{\rm{agb}}}$ value is a constant. Some results are
presented in Fig.13, Fig.14, and Fig.15. The sequences of
$\dot{M_{\rm{agb}}}$=2$\times 10^{-5}M_{\odot}$ yr$^{-1}$ are in Fig.13,
those of $\dot{M_{\rm{agb}}}$=3$\times 10^{-5}M_{\odot}$ yr$^{-1}$ are
in Fig.14, and those of $\dot{M_{\rm{agb}}}$=4$\times 10^{-5}M_{\odot}$
yr$^{-1}$ are in Fig.15.  In these calculations, we set $\alpha$=0.5,
$C_{\rm{s}}$=10 km s$^{-1}$ and examine the range of $v$=500-10000 km
s$^{-1}$ at the steps of 500 km s$^{-1}$. The three cases of
$V_{\rm{agb}}$=10,15,20 km s$^{-1}$ are examined.  In those figures, we
plot only the models which satisfy the $R_{\rm{n}}$-$M_{\rm{n}}$
relation if the lines turn around the point of
$\rm{log}(R_{\rm{n}})\sim$ -1.0 and $\rm{log}(M_{\rm{n}})\sim$ -1.0 in
$\rm{log}(R_{\rm{n}})$-$\rm{log}(M_{\rm{n}})$ plane.  The
$R_{\rm{n}}$-$M_{\rm{n}}$ relations of the selected models are shown in
Fig.16, Fig.17, and Fig.18.
 
{}From Figs.13, 14, and 15, when $\dot{M_{\rm{agb}}}$=const, we can
re-product only the region of $2V_{\rm{exp}} \leq$ 70 km s$^{-1}$. The
models which are able to trace the realm of larger expansion velocity do
not match the $R_{\rm{n}}$-$M_{\rm{n}}$ data.

Comparing time-depending mass loss models (Fig.6) with steady mass loss
models (Figs.13, 14, 15), one can know how important for the evolution
of PNe to take account of the variation of AGB mass-loss rate.  In other
words, the radial density distribution of ambient matter around PN,
$\rho(r)$, is not in proportion to $r^{-2}$. It needs to fall suddenly
at $r\sim$0.1pc due to the pulsation history during AGB phase.
Furthermore, this indicates the difference between the dynamical age
and evolutionary age of PNe since the expansion law is affected 
by the density profile around the proto-PNe. We expect 
the age-discrepancy problem of PNe (Mellema 1994) may be partially
resolved if the origin of the velocity-radius relation is resolved.

By the way, the variety of $v$ contributes greatly to the explanation of
data distribution in $2V_{\rm{exp}}$-$R_{\rm{n}}$ plane.  In some
models, $v$ has a very large value up to 10000 km s$^{-1}$ at $t=0$.
However, $v$ may not attain such a large speed in the theory of
line-driven wind. We deals with this problem in \S\S 4.4 again.
 
\subsection{Observational implications}
In $2V_{\rm{exp}}$-$R_{\rm{n}}$ plane, most of observational data exist in 
the range of $2V_{\rm{exp}}$=40-80 km s$^{-1}$ and $R_{\rm{n}}$=0.05-0.15pc. 
We present by our calculations that these data are in the realm of 
$t$=2000-4000 yr(cf. Fig.6 in Sect.2) and it coincides with the fact that 
PN is an object between the AGB star and white dwarf. 
Our results suggest the reason why only few data is in 
the smaller radius region. PNe evolve very quickly across this 
region, and then it is difficult for us to observe. 
In the larger radius region, it is also difficult to observe since PNe 
expand too large and diffuse. 
By a more highly accurate survey of PNe, in future, 
we will find that there are a lot of faint PNe and some of them have
larger radius among them. 
  
\subsection{Velocity problem}

As stated in subsection 4.2, we have a time gap problem concerning the
fast wind velocity from the central star of PN.  Along the line-driven
wind theory, it takes about 8000 yr until $v$ reaches 10000 km s$^{-1}$,
while in our calculation it needs to be attained at $t$=0.  However we
will solve this problem in two different simple ways: (1) accretion jet
and (2) compactness of the central star.  We shall examine the case of
(1).  In the line-driven wind it is assumed the PN field is spherical
symmetry.  The assumption is not so realistic and many PNe have an
asymmetric jet due to its magnetic field.  In this way, the gas of PN is
accelerated so quickly that our supposition for $v$ is good.  By this
way of thinking, we may say that the reason why few PNe are in the
larger expansion velocity and smaller radius region in their early phase
($<$2000 yr) is that the total number of asymmetrically accelerated PNe
is small in their early evolution.

The other explanation of the case of (2) is rather simple, although the
acceleration mechanism can not be specified. If the central star has
very small radius ($\sim$1000 km), its gravity at the surface of the
star is so large. To escape from the central star gravity field, a
matter must have a very large velocity.  In fact, a radius of typical
white dwarf is in the range of 1000-10000 km.  It is expected that the
acceleration mechanism of PNe will be proved by future observations.
 
\section{Summary}

Constructing the modified ``two-wind'' model, we find that the
observational data are reproduced adequately if the radial density of
the AGB matter has a sudden fall at about 0.1pc from the central star.
It corresponds the AGB star pulsates and its mass loss rate
increases. When we set the start point of evolutionary track of PN is
$t$=0 and the velocity of the AGB matter ($V_{\rm{agb}}$) is 10 km
s$^{-1}$, we estimate $\dot{M_{\rm{agb}}}$=$10^{-6.4}M_{\odot}$ at
$t$=$-10^{4.5}$yr to $\dot{M_{\rm{agb}}}$=$10^{-5}M_{\odot}$ at
$t$=$-10^{3.8}$yr.  It means that the features of each PN considerably
depend on its progenitor's mass loss history, and by only including the
effect of time-dependent mass loss rate of the AGB star we can explain
the observed data in the context of fluid dynamics.  Also our model
gives an observational suggestion that the PNe with larger Vexp ($\geq$
25km s$^{-1}$) and smaller radius($\leq$ 0.1pc) are evolving so quickly
that it is difficult to observe them.  However, to completely account
the observational trend, we should investigate the acceleration
mechanism of the fast wind since as long as spherical symmetry model is
adopted, we need the long time-scale to reach at 10000 km s$^{-1}$.  In
order to prove the acceleration mechanism of the shell of PN, we have to
do a more accurate survey of PNe because observations of only very young
PNe make us know the mechanism.  We expect that such observations will
be done in the near future.

\bigskip

We are grateful to the referee. His/Her comments help us to improve
the paper. 
This work is supported by the Grant-in-Aid for the 21st Century COE 
"Center for Diversity and Universality in Physics" 
from the Ministry of Education, Culture, Sports, 
Science and Technology (MEXT) of Japan,
and partially supported by 
MEXT of Japan for 16740110 and 18026005.

\section*{References}
\small

\re
Bl$\ddot{\rm{o}}$cker, T. 1995, A\&A, 297, 727

\re 
Bl$\ddot{\rm{o}}$cker, T. 1995, A\&A, 299, 755

\re
Bohuski, T. J. \& Smith, M. G. 1974, ApJ, 193, 197

\re
Frank, A. 1994, AJ, 107, 261

\re
Kwok, S. 1978, ApJ, 225, 107

\re
Kwok, S. 1982, ApJ, 258, 280

\re
Kwok, S., Purton, C. R., \& Fitzgerald, P. M. 1978, ApJ, 219, 125

\re
Martin, H \& Sch$\ddot{\rm{o}}$nberner, D. 1991, A\&A, 248, 590

\re
Mellema, G., 1994, A\&A, 290, 915

\re
Perinotto, M., Sch$\ddot{\rm{o}}$nberner, D., Steffen, M.,
\& Calonaci, C., 2004,  A\&A, 414, 993

\re
Renzini, A. 1979, Proceedings of the Fourth European Regional Meeting in Astronomy (Dordrecht, D. Reidel), 155

\re
Robinson, G. J., Reay, N. Y., \& Atherton, P. D. 1982, MNRAS, 199, 649

\re
Sabbadin, F., Benetti, S., Cappellaro, E., Ragazzoni, R., \& Turatto, M.,
 2005, A\&A, 436, 549

\re
Sabbadin, F., Gratton, R. G., Bianchini, A., \& Ortolani, S. 1984, A\&A, 136, 181 (SGBO)

\re
Sabbadin, F. \& Hamzaoglu, E. 1982, A\&A, 110, 105

\re
Schmidt-Voigt, M. \& K$\ddot{\rm{o}}$ppen, J., 1987, A\&A, 174, 211

\re
Schmidt-Voigt, M. \& K$\ddot{\rm{o}}$ppen, J., 1987, A\&A, 174, 223

\re
Sch$\ddot{\rm{o}}$nberner, D., Jacob, R., \& Steffan, M.,
2005, A\&A, 441, 573

\re
Sch$\ddot{\rm{o}}$nberner, D. Jacob, R., Steffan, M.,
Perinotto, M., Corradi, R. L. M., \& Acker, A., 2005, A\&A 431, 963

\re
Smith, Dean Francis. 1969, PhDT, 6

\end{document}